# A novel method to analyze pattern shifts in rainfall using cluster analysis and probability models

Abhishek Singh, Aaditya Jadhav, Abha Goyal, Jesma V and Vyshna I C
Dept. of Agricultural Engineering, Institute of Agricultural Sciences
Banaras Hindu University

**Abstract**: One of the prominent challenges being faced by agricultural sciences is the onset of climate change which is adversely affecting every aspect of cropping. Modelling of climate change at macro level have been carried out at large scale and there is ample amount of research publications available for that. But at micro level like at state level or district level there are lesser studies. District level studies can help in preparing specific plans for the mitigation of adverse effects of climate change at local level. An attempt has been made in this paper to model the monthly rainfall of Varanasi district of the state of Uttar Pradesh with the help of probability models. Firstly, the pattern of the climate change over 122 years has been unveiled by using exploratory analysis and using multivariate techniques like cluster analysis and then probability models have been fitted for selected months.

**Introduction:** Agriculture is the most prominent domain of all the human endeavors which is acutely affected by the climate change affecting the human civilization. Particularly in this century the adverse effects have been highly visible with extreme climate events like unusually high temperatures, heat waves, droughts, forest fires, extreme rainfall, floods, cloud bursts and cyclones occurring more frequently than earlier. Mitigation of climate change seems to be a distant dream because of the changing policies of the world's powerful countries like USA. Though IPCC established in the year 1988, by the United Nations Environment Programme (UN Environment) and the World Meteorological Organization (WMO), is doing a commendable work for mitigating the effects of climate change through its regular assessments of the climate risks. The world is also moving slowly towards the goal and all the efforts are directed to limit the global warming by less than $1.5^\circ$ C up to the end of this century.

While the governments are busy in framing policies to mitigate global warming, it is also pertinent to enhance tools and techniques to adapt to changes which have already taken place. Particularly at micro geographical levels like at district or block level adaption, this will greatly help in coping with the adverse effects of climate change. This involves first ascertaining the climate change patterns at local level and then chalking out specific plans.

Whenever farmers are planning to sow a crop they take into account the date of sowing and the weather conditions which usually prevail in that area and they plan according to the past pattern of climate conditions. They also account for possible rainfall which will reduce the requirement of paid irrigation and thus saving them some money. Also the date of harvesting is also taken into account. A changed pattern of rainfall fails all their planning which results in financial and social stress. Therefore, quantifying the changing pattern becomes a pertinent and urgent task. Statistical methods like clustering and probability models can help the stakeholders to mitigate this problem to some level.

Statistical techniques like cluster analysis and probability models are used to identify the shifts in the pattern and later on quantifying it with the purpose of forecasting using probability models. Both of the techniques are very common and used in diverse area of study (Al Mamoon A & Rahman A ,2017; Ashok Rai *et.al.*2019; Pal R & Pani P,2016;). But cluster analysis which is a multivariate technique is rarely used to ascertain the pattern of rainfall in year. This technique is generally used to from natural groups in multivariate data with the help of distances. Here it has been used to form groups or clusters of years from 1901 to 2022 on the basis of the distribution of rainfall over the months in the year. The years having the same pattern of rainfall over the months has been clustered together. The obtained clusters clearly show that the pattern of rainfall in different months of year has changed. In order to ascertain the distribution shifts in rainfall patterns in a year, decadal averages of the monthly rainfall were taken and clustering was carried out to group different decades on the basis of the patterns.

Probability distributions are used to fit rainfall data in diverse fields like hydrology, meteorology (Sharma C & Ojha CSP,2019; Teodoro PE *et.al.*2016; Deniz Ozonur *et.al.*2021; Teodoro PE *et.al.* 2015; Blain GC & Meschiatti MC,2014; Parida BP,1999) etc.in order to ascertain not only the total rainfall but the distribution of the rainfall in a season, because the yield of a crop is not affected by the total rainfall but by the pattern of the rainfall over a season. These distributions can be used to quite accurately estimate the amount of precipitation and its pattern. This becomes a pertinent task for deciding dry and wet seasons so that appropriate planning can be done to manage water resources for agriculture.

**Material and Methods**

**Data Consideration**

Monthly Rainfall data of Varanasi was considered for the study. Data from 1901 to 2022 was collected. In total 1464 data points were there for analysis. Total Monthly data for 122 years was considered. The source of the data was from the website: www.indiawaterportal.org.

Grouping of individuals when there is only one variable measured on each individual is relatively easy and can be done by making classes and forming frequency distributions etc. But when large number of variables is measured on each individual of the population forming groups becomes a complex task. Here in comes the role of multivariate statistics in which by using the technique of cluster analysis individuals can be grouped when large number of variables are measured on each individual. In cluster analysis the patterns in a data set are explored by grouping the observations in to clusters. The prime aim of clustering is to form optimal groups such that observations in a group are similar within a group but the groups are dissimilar to each other's. The methods used for clustering are Hierarchical clustering and partitioning. In hierarchical clustering we typically start with n clusters, one for each observation, and end with a single cluster containing all n observations. At each step, an observation or a cluster of observations is absorbed into another cluster. We can also reverse this process, that is, start with a single cluster containing all n observations and end with n clusters of a single item each. In partitioning, we simply divide the observations into g clusters. This can be done by starting with an initial partitioning or with cluster centers and then reallocating the observations according to some optimality criterion. Several computer algorithms have been developed to form clusters. Hierarchical methods have been used in the present study which is of two types- agglomerative and divisive. In this process, the number of clusters becomes smaller and smaller and the clusters themselves grow larger. We begin with n clusters (individual items) and end with one single cluster containing the entire data set. An alternative approach, called the divisive method, starts with a single cluster containing all n items and partitions a cluster into two clusters at each step The end result of the divisive approach is n clusters of one item each. Agglomerative methods are more commonly used.

For cluster analysis SPSS package was used and for fitting of the probability models, different packages of R like Fitdistr,Fitdistrplus,univariateML which provides a model select function that automatically tests and fit many different distributions and then selects the best fit based on AIC (default), BIC, or log likelihood. Different probability models that were fitted are- beta,betapr,Cauchy ,exp ,gamma, ged, gumbel, invgamma, invgauss, invweibull, laplace,lgamma, llogis, lnorm,logis,logitnorm,lomax, naka, norm, pareto,power, Rayleigh, sged,snorm,sstd,std, ,unif,Weibull.

**Results and Discussion**

Monthly Distribution of the Rainfall: Monthly distribution of the rainfall over the years is shown in the figure 1. It shows that the maximum rainfall in Varanasi during the year is in the months of June, July, August and September with the peaks visible in the month of July, August and September. Almost 70 percent of the total rainfall in the year falls in the month of July, August and September.

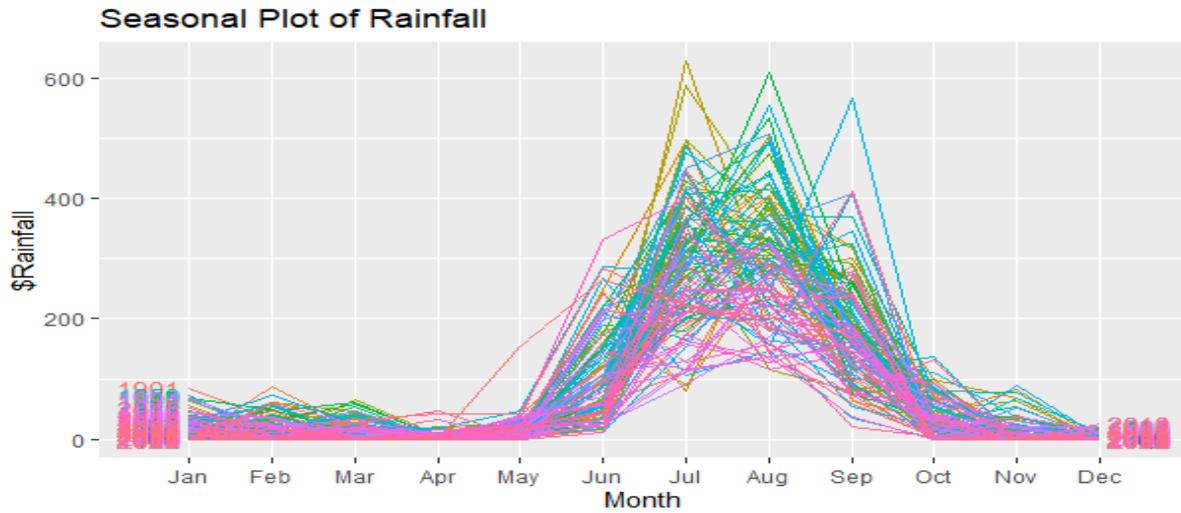

Fig. 1: Monthly distribution of rainfall in Varanasi over the years

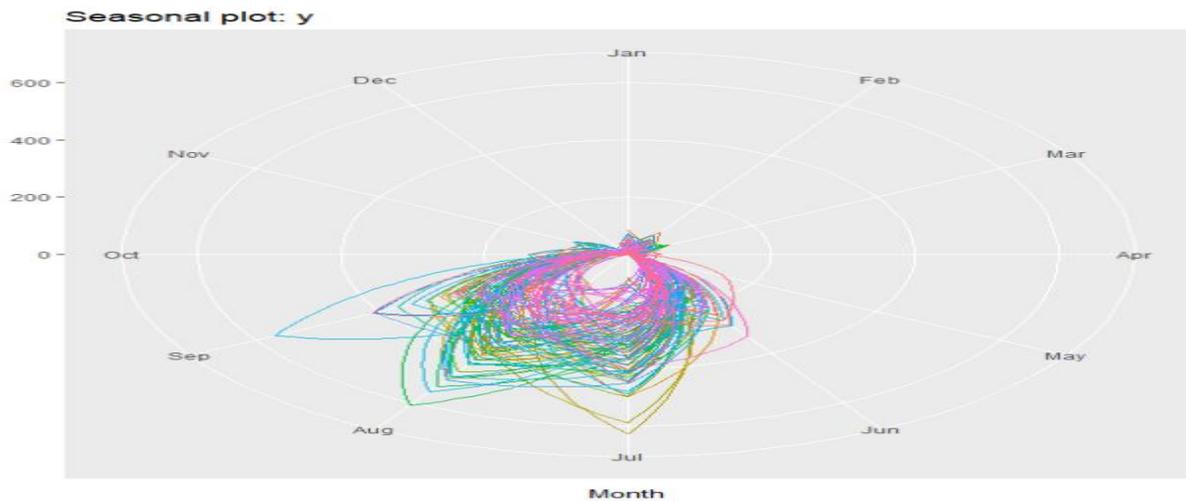

Fig.2 : Polar plot for showing the distribution of monthly rainfall of Varanasi District

The polar plot of the monthly rainfall distribution (Fig. 2) also indicates the same pattern that most of the annual rainfall occurs in the monsoon months of June, July, August and September.

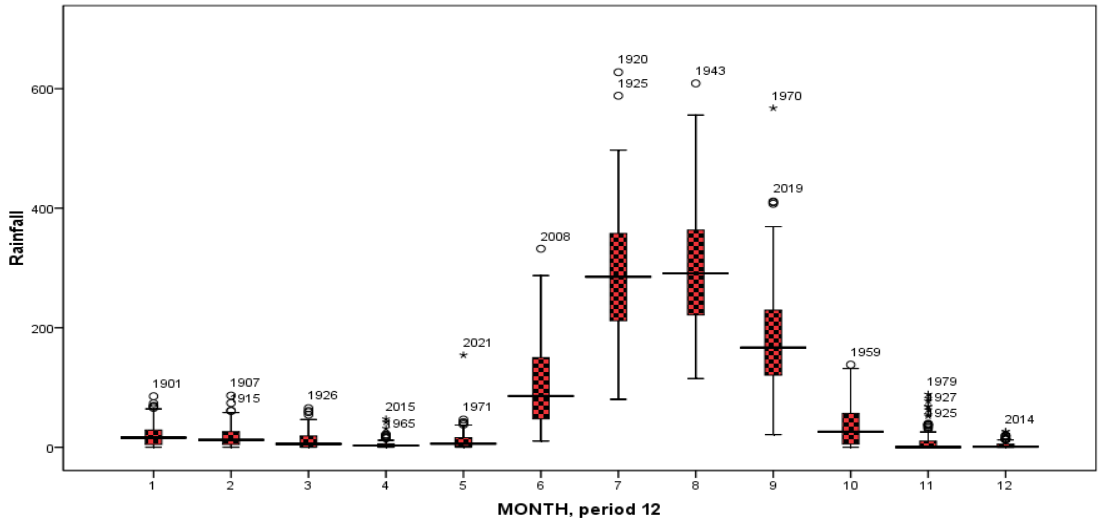

Fig. 3: Box and whisker plots to depict monthly distributions over the years

The monthly variation in the different months is depicted in Box and Whisker plots which is five point depiction of the data showing a datasets Quartile1, Quartile 2 and Quartile 3 as well as Upper adjacent values and Lower adjacent value. The variation in the data is depicted by the length of the Box which shows the variation in the middle 50 % of the observations. The central tendency is shown by the horizontal line in the box which indicates the value of the median. The data depicted in the fig 3 indicates that the maximum rainfall occurs in the month of August in Varanasi closely followed by the month of July then September followed by June and October in that order. The minimum rainfall is in the month of November and December. If we look at the length of the boxes which are indicative of the variation in the rainfall of a particular month it can be deducted that the maximum variation occurs in the month of August and July followed by that September, June and October. The lowest variation in the rainfall is in the months of December and April. If we look at the outlier values of different months it clearly indicates that some extreme events have happen in these months with respect to the rainfall. Out of these extreme values some of them have happened in this century like for the month of April it is the Year 2015, for May it is the year 2021, for June it is 2008, for July it is 1920 and 1925, for August it is 2019 and 1970. It clearly indicates that there is a shift in the pattern of rainfall, Instead of maximum rainfall in the months of July and August months of May , June and October are also witnessing some unusually high rainfall which is indicative of pattern change in the rainfall.

Descriptive statistics for the monthly rainfall data

**Table :1 - Descriptive statistics of the monthly rainfall in Varanasi District**

| Summary Statistics | Mean | Median | Mode | Standard Deviation | Kurtosis | Skewness | Range | Minimum | Maximum |
|---|---|---|---|---|---|---|---|---|---|
| January | 19.33 | 16.12 | 0.2 | 18.52 | 1.68 | 1.34 | 85.36 | 0 | 85.36 |
| February | 17.92 | 12.43 | 0 | 17.98 | 1.68 | 1.39 | 86.54 | 0 | 86.54 |
| March | 12.16 | 5.55 | 0 | 15.34 | 1.75 | 1.52 | 65.31 | 0 | 65.31 |
| April | **5.04** | 2.9 | 0 | **7.44** | 13.73 | 3.28 | 47.46 | 0 | 47.46 |
| May | 11.78 | 6.12 | 0 | 17.32 | 37.4 | 4.96 | 154.37 | 0 | 154.37 |
| June | 102.08 | 85.85 | 94.92 | 71.8 | 0.34 | 1 | 321.89 | 10.35 | 332.23 |
| July | **285.84** | 285.32 | 221.48 | **107.18** | 0.18 | 0.42 | 547.37 | 80.24 | **627.61** |
| August | **298.39** | 291.15 | 184.57 | 105.83 | -0.16 | 0.44 | 493.94 | 114.95 | **608.89** |
| September | 179.41 | 166.86 | 73.83 | 87.45 | 2.79 | 1.17 | **546.54** | 21.09 | 567.63 |
| October | 34.02 | 26.02 | 5.27 | 32.57 | 0.28 | 1 | 138.24 | 0 | 138.24 |
| November | 10.06 | 0.37 | 0 | 19.03 | 5.43 | 2.39 | 88.79 | 0 | 88.79 |
| December | **3.85** | 0.91 | 0 | **5.85** | 3.45 | 1.92 | 26.37 | 0 | 26.37 |

In the table 1: the summary statistics of the months over the years have been shown. It indicates that the maximum average rainfall occurs in the month of August (298.39), closely followed by the month of July (285.84) and the minimum average rainfall was observed in the month of December (3.85) followed by April (5.04). The maximum variation was observed in the month of July (107.18), followed by the month of August (105.83) and the minimum variation was observed in the month of December (5.85) followed by the month of April (7.44). Maximum range of rainfall was obtained in the July (547.37) and September (546.54). During these years the maximum rainfall is obtained in the month of July (627.61) followed by the month of August (608.89).

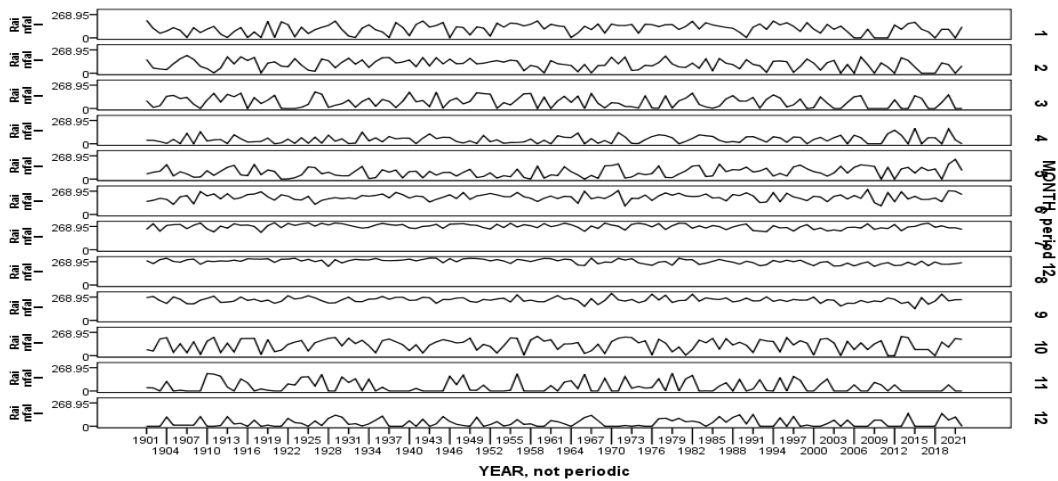

Fig.4 : - Monthly trends of rainfall over the years (1901-2018)

Trends of monthly rainfall over the years have been shown in the fig. 4 which clearly shows that there are no prominent long term trends in the rainfall of different months. Cyclic patterns are observed in the rainfall of every month but the length of the cycles are not constant and changes frequently. The curves of the months July, August and September are touching the upper boundaries of the graph indicating that maximum rainfall occurs in these months of the year.

Annual distribution of the total rainfall is depicted in the fig. 5 which indicates that up to the year 1985, the trend of the rainfall is same but after that a negative trend in the total rainfall is observed. In fact, two of the years with minimum total rainfall are observed in the first decade of this century i.e. 2004 (495.71) and 2009 (516.79) while two years of total annual rainfall were observed in the years 1925 (1455.269) and 1956 (1461.409). This clearly points out the total rainfall in the Varanasi district is showing a downward movement.

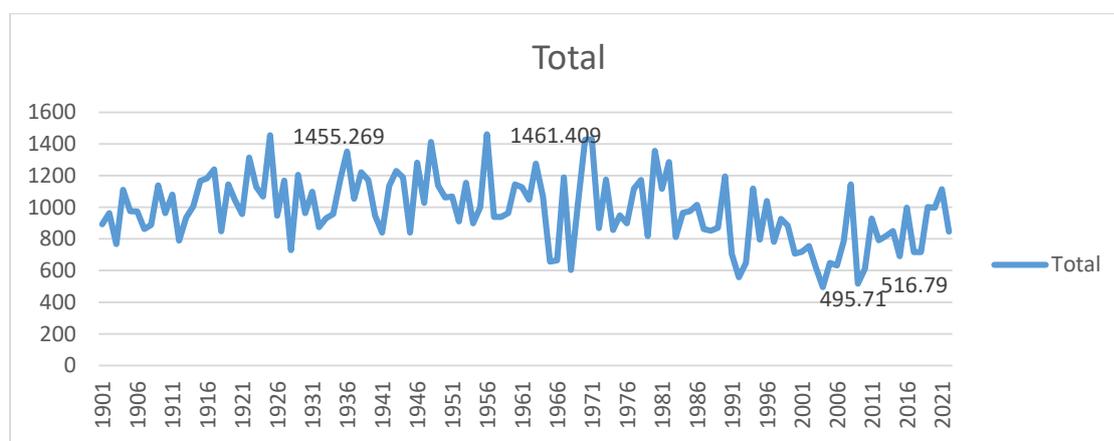

Fig. 5: Trend of Annual Rainfall over the years (1901-2021) in Varanasi

Cluster analysis is multivariate technique to identify natural groupings among items characterized by multiple factors. In this paper the technique has been used to form clusters of years having similar pattern of rainfall distribution over the months in a year. Rainfall in different months have been considered as different variables and then they are used for forming clusters. The results have been shown in the figure 6. The prominent clusters have been shown with the help of circles. It can be seen that the rainfall distribution in the year 2016 was similar to that of the year 1947. Similarly, 1902 and 2012, 1974 and 2017,1987 and 2002,1984 and 2011, 1945 and 2022 were having similarly monthly distribution of rainfall. But at the same time the monthly distribution of rainfall in most of the years of this century is different from other years. The years 2003 and 2008, 2001 and 2005, 2010 and 2007, 2006 and 2018, 2020 and 2021,2004 and 2013 and 2004 are all having similar monthly distributions.

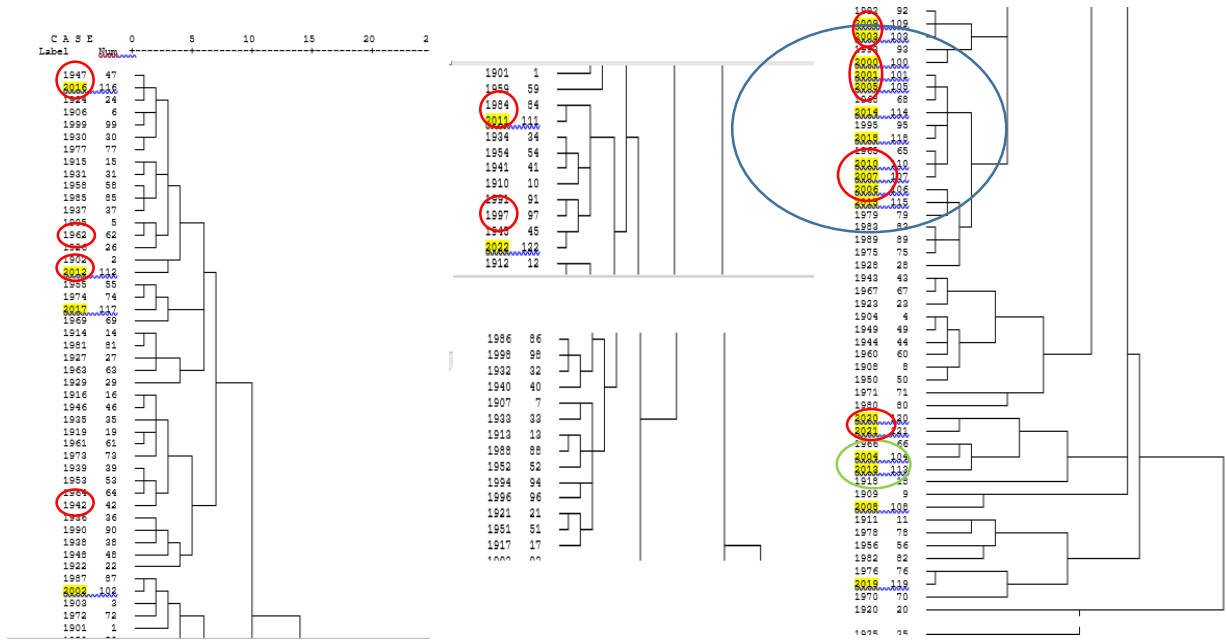

Fig. 6 :- Clusters of years in the terms of the monthly distribution of rainfall

Decadal Distribution of the Rainfall : In order to identify and quantify the changing patterns, data for different months were averaged over a decade and decadal pattern of the monthly rainfall were obtained. The whole data was divided in to thirteen decades. The pattern is displayed in the fig. 7. This clearly indicates that the decadal averages of the monthly rainfall have generally reduced in the most recent decades.

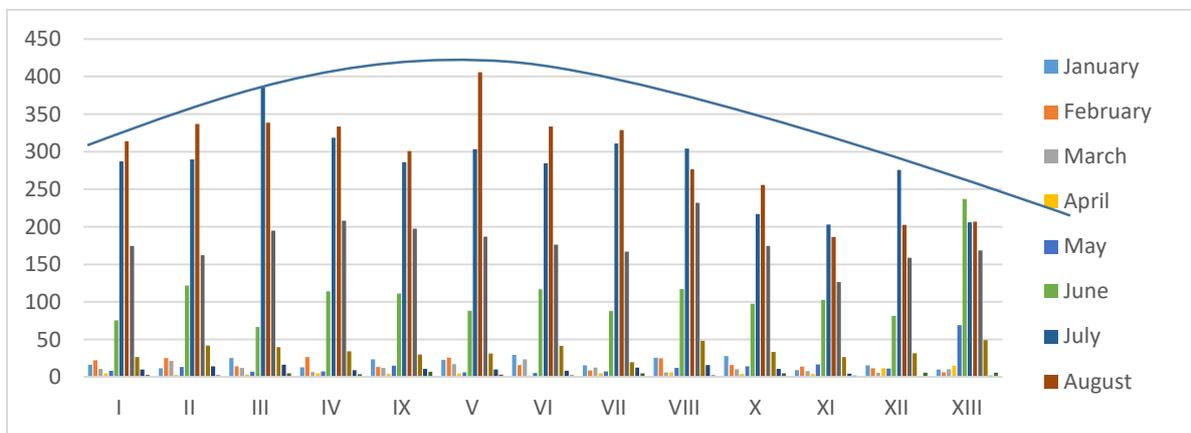

Fig. 7 : Decadal monthly Pattern of Rainfall distribution of Varanasi

Grouping of these decades were also performed to identify the changing pattern of monthly rainfall in different decades (Fig.8). It is visible in the dendrogram that the decades 10,12 and 11 formed a different cluster based on the rainfall patterns in different months, while the last decade of the study is having a vastly different distribution of rainfall in comparison to all the other decades. This is indicative of the drastic change in the monthly rainfall pattern due to the climate change.

Decadal Cluster

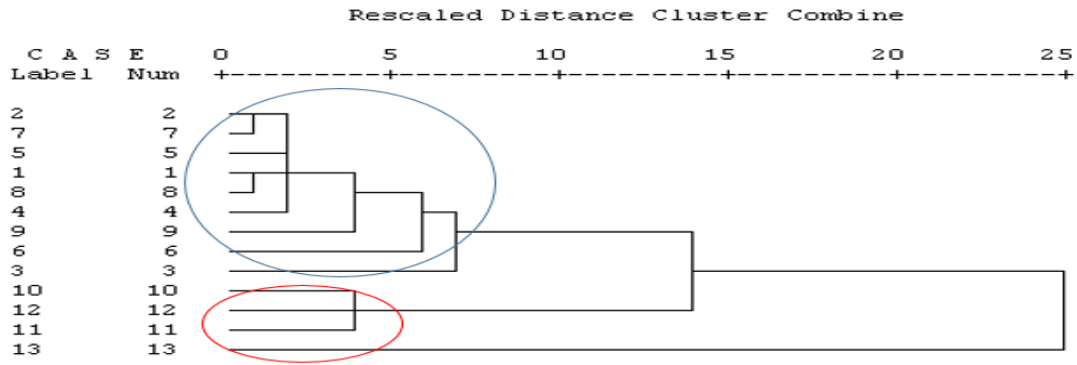

Fig. 8 :- Cluster of Decadal Rainfall Distribution in Varanasi

After ascertaining the fact that the rainfall distribution in different months have drastically changed, different probabilistic models were fitted to the rainfall distribution of each month. Cullen and Frey graph, Q-Q plots, P-P plots etc. were used to check the model fit (Fig. 9 & 10).

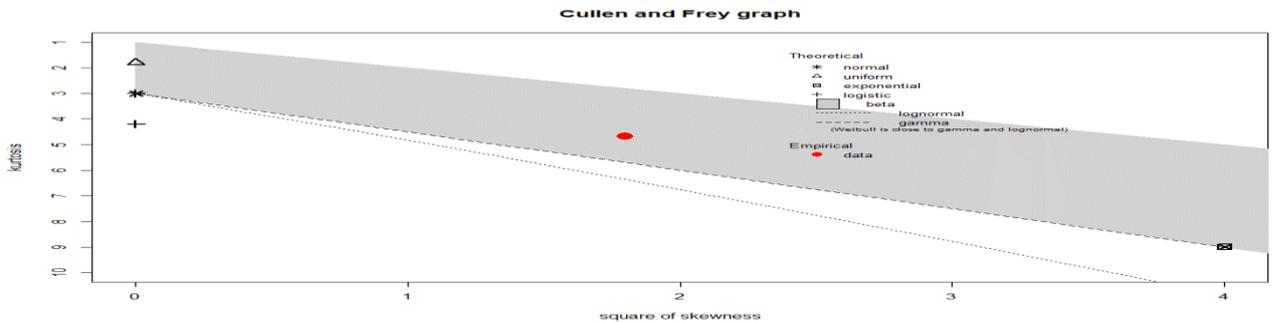

Fig. 10: Cullen and Frey Graph

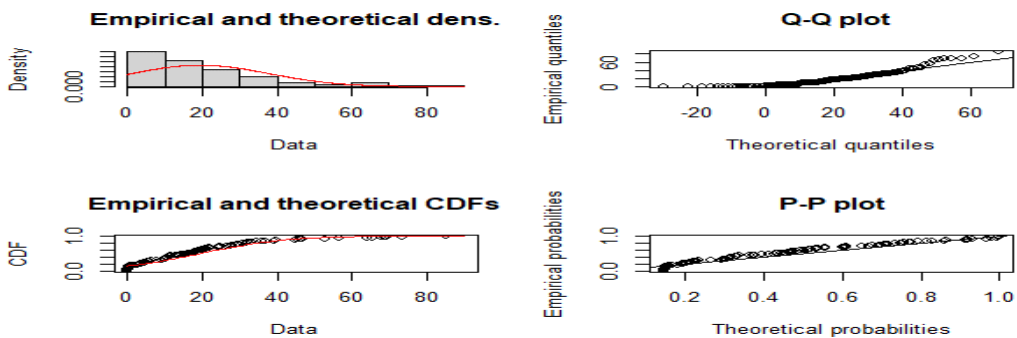

Fig. 11: Different graphical measures to check the fitted models

Some of the results of different fitted models have been displayed in the fig.11. For the months of January and February exponential model, for the month of July Nakagami model , for the month of August Gamma Model and for the month of September Gumbel model were found to be the best fitted probability distributions.

| | | |
|---|---|---|
| 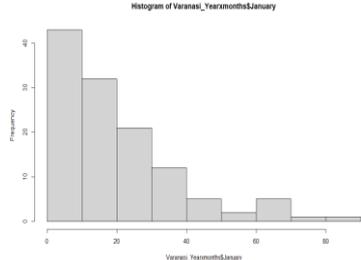 | Exponential model<br><br>Estimates:<br> rate  0.05172944<br>Log-likelihood:<br>-483.3308 | 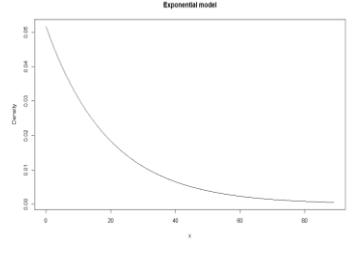 |
| 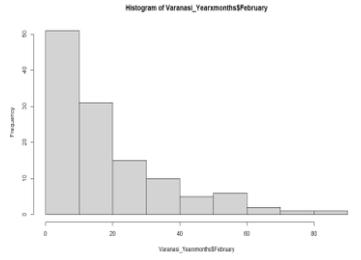 | Exponential model<br><br>Estimates:<br>Rate 0.05581314<br><br>Log-likelihood:<br>-474.061 | 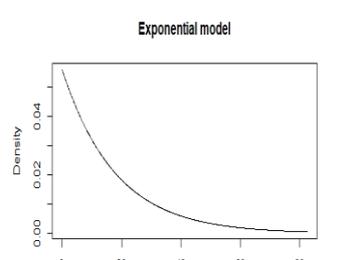 |
| 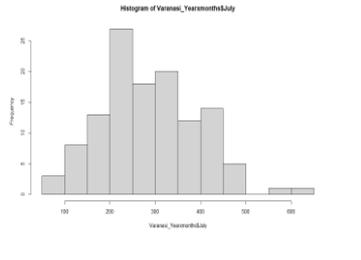 | Nakagami model<br><br>Estimates:<br>Shape 1.900265<br>scale 93098.633771<br><br>Log-likelihood:<br>-740.4096 | 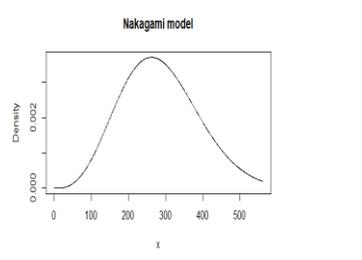 |
| 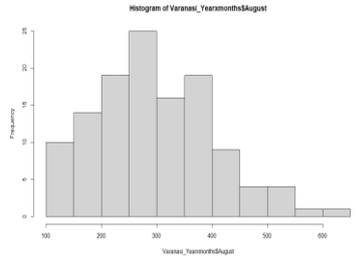 | Gamma model<br><br>Estimates:<br>shape 7.72732458<br>rate 0.02589706<br>Log-likelihood:<br>-738.1478 | 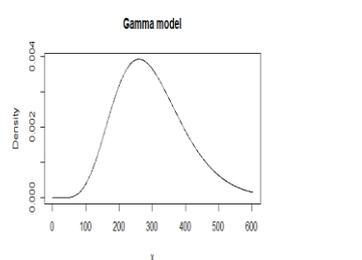 |
| 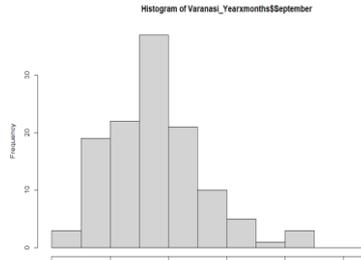 | Gumbel model<br><br>Estimates:<br> mu        140.07695<br><br>sigma    69.19052<br><br>Log-likelihood:<br> -708.2596 | 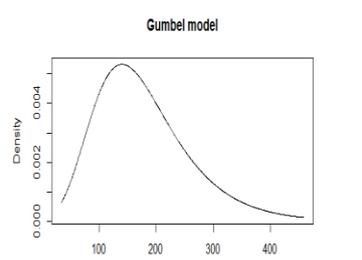 |

Fig. 9 : Observed and Fitted probability distribution in different months

**Conclusions**

Cluster analysis was used to identify the clusters of different years having similar monthly rainfall patterns within a year. It was observed in the study that the years of this century formed different clusters, indicating that the rainfall patterns of different months have considerably changed from the earlier years. Decades were also grouped on the basis of average monthly rainfalls and it was observed that the 13$^{th}$ decade of the study was having entirely different average monthly rainfall patterns. Thus cluster analysis was used to identify changing rainfall patterns and the study indicated that climate change has indeed change the monthly distribution of rainfall within a year. Further in this study, several probability distributions GUM, LN2P, LN3P, W2P, W3P, G2P, G3P were applied in order to determine appropriate probability distribution for modelling monthly rainfall data in the Varanasi District of Uttar Pradesh. Maximum likelihood method was used to estimate parameters of the examined distributions. Several goodness of fit tests, namely chi-squared and Kolmogorov–Smirnov tests, were used to decide suitability of the probability distributions. In terms of modelling monthly rainfall data, the goodness of fit results indicate that the best distribution can differ from month to month and there is no single distribution providing the best fit for all of the months. The results presented in this study can be used to better understand and model more accurately rainfall in the Varanasi District.